\documentclass[twocolumn]{aa}
\usepackage{graphicx}
\usepackage{natbib}
\begin{document}
\title{Detection and analysis of a new INTEGRAL hard X-ray transient, IGR J17285-2922}

\titlerunning{IGR J17285-2922}

\author{E.~J. Barlow\inst{1}
\and A.~J. Bird\inst{1}
\and D.~J. Clark\inst{1}
\and R. Cornelisse\inst{1}
\and A.~J. Dean\inst{1}
\and A.~B. Hill\inst{1}
\and L. Moran\inst{1}
\and V. Sguera\inst{1}
\and S.~E. Shaw\inst{1,2}
\and D.~R. Willis\inst{1}
\and F. Capitanio\inst{3}
\and M. Del Santo\inst{3}
\and L. Bassani\inst{4}
}
\institute{School of Physics and Astronomy, University of Southampton, Highfield, SO17 1BJ, UK
\and
{\it INTEGRAL} Science Data Centre, CH-1290 Versoix, Switzerland
\and
IASF/CNR, Via del Fosso del Cavaliere 100, 00133 Roma, Italy
\and
IASF/CNR Section of Bologna, Via P.~Gobetti 101, 40129 Bologna, Italy
}

\offprints{Liz Barlow, ejb@astro.soton.ac.uk}

\abstract{We present the transient nature of INTEGRAL source IGR~J17285--2922, identified from a single period of activity during an IBIS/ISGRI Galactic Centre Deep Exposure in September 2003.  The source has a maximum detection significance of 14$\sigma$ in the 20--100~keV energy range and exhibits a flux of 6.5~mCrab before it moves out of the ISGRI field of view.  The source is visible to at least 150~keV and its spectrum can be fit with a power law slope of $\Gamma=2.1 \pm0.17$; a more physical model could not be fit due to poor statistics.  Detected characteristics are consistent with the source being a Galactic low-mass X-ray binary harbouring a black hole or neutron star.

   }

\maketitle

%

\section{Introduction}
In the two years since {\it INTEGRAL} has been in orbit \citep{INTEGRAL}, over 40 new high energy ($\geq$20~keV) sources have been discovered with the gamma-ray imager, IBIS/ISGRI \citep{Ubertini,ISGRI}.  Several of these new sources have been classified e.g. IGR~J16318--4848 \citep{4848}, but the majority remain poorly understood.  The IBIS/ISGRI Survey Catalogue presented the detection of 123 sources in the energy range 20-100~keV, of which 28 were unidentified at the time \citep{catalog}.  The survey utilises Core Programme data that consists of deep exposures of the Galactic Centre region and regular scans along the Galactic Plane.    

As a result of the exposure bias on the Galactic Centre and Plane, the majority of the catalogue sources are X-ray binaries, classified according to the mass of the donor star.  Low-mass binaries (LMXBs) are generally located in the Galactic Bulge and high-mass binaries (HMXBs) in the Galactic Plane following the expected distributions of their parent stellar populations; see Psaltis \citep{population} for a review. Therefore, we can expect most of the unidentified sources to be X-ray binaries, including systems containing a black hole and possibly a new class of highly absorbed objects characterised by an absence of soft X-ray emission \citep{absorbed}.  The IBIS/ISGRI survey provides a unique opportunity to study the long term behaviour of persistent hard X-ray/soft gamma-ray sources and to identify faint transient objects that would have otherwise remained undiscovered.

IGR~J17285--2922 was first discovered from investigation of the IBIS/ISGRI survey mosaics \citep{ATEL}, but was not included in the first IBIS/ISGRI catalogue as the source's detection significance (based on mean survey source flux) was below the 6$\sigma$ threshold set to minimise the chance of false detections \citep[from a systematic logN/logS analysis,][]{catalog}.  However, by re-analysing the individual sets of GCDE observations it can be shown that the source is transient, with a maximum 20-40~keV detection of 11$\sigma$ at ($\alpha$,$\delta$)=($17^{h}28^{m}41^{s}\, -29^{\circ}22\arcmin56\arcsec\,)\pm 2\arcmin\!\ $).  The source is near to the location of the Galactic Centre at (${\it l}$,${\it b}$)=($357^{\circ}\!.64 , +2^{\circ}\!.88$) and within 1.5$^\circ$ of the bright X-ray source, 4U~1724--307, associated with the globular cluster Terzan 2.  A search of existing source catalogues has yielded no known X-ray counterparts within the 2' error circle.  It has been found that $\sim$60\% of IBIS/ISGRI survey sources have {\it ROSAT} counterparts \citep{rosat} but no such association has been found for this source.  A search at longer wavelengths is impractical at present due to the relatively large error circle and crowded region of the sky.  We present analysis of the timing and spectral characteristics of IGR~J17285--2922 and discuss its possible nature.

\begin{table*}[t]
\begin{center}
\begin{tabular}{l l l l l}
\hline 
\hline GCDE & Obs Date & Rev & IJD & t$_{eff}$ (ksec)\\
\hline
A & 27/02/03 - 04/05/03 & 46-67   & 1153.9 - 1219.6 & 471\\
B & 08/08/03 - 20/08/03 & 100-1,103  & 1315.4 - 1327.3 & 96\\
C & 25/09/03 - 19/10/03 & 116-123  & 1363.3 - 1387.1 & 191\\
D & 12/02/04 - 24/04/04 & 163-186  & 1503.9 - 1575.6 & 550\\
\hline
\end{tabular}
\caption{\label{gcde} {Overview of the Galactic Centre Deep Exposure (GCDE) observations, giving the date, revolution numbers, time in {\it INTEGRAL} Julian Date and on source effective exposure, t$_{eff}$.  IGR~J17285--2922 was detected in revolutions 118-121 in Group C only.  {\it INTEGRAL} Julian Date (IJD) = MJD - 51544.}}
\end{center}
\end{table*}

\begin{figure}[h]
\begin{center}
\includegraphics[totalheight=5.0cm]{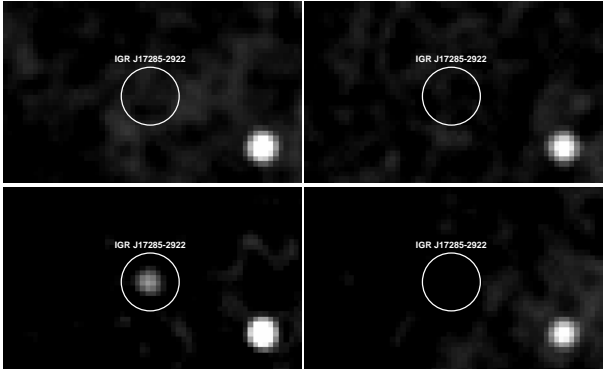}
\caption{\label{GCDE} {Significance mosaics (20--100~keV energy band) showing the region around IGR~J17285--2922 (indicated by circle).  Top left: Mar/Apr 2003 (A), top right: Aug 2003 (B), bottom left: Sep/Oct 2003 (C), bottom right: Feb--Apr 2004 (D).  The bright source in the bottom right of the frame is the Terzan 2 X-ray source. For mosaic exposures see Table~\ref{gcde}.}}
\end{center}
\end{figure}

\section{Observations and Data Analysis}
The Galactic Centre Deep Exposure (GCDE) is performed at $\sim$6 month intervals and consists of $\sim$1 month of quasi-continuous observations.  Table~\ref{gcde} summarises GCDE observations taken between February 2003 and April 2004.  Each orbit (revolution) of GCDE data is divided into $\sim$2000s long science windows (ScW).  Data was processed using the Offline Standard Analysis, OSA v4.1 software \citep{osa} available from ISDC \citep{ISDC}.  ScW images can be combined together to form mosaics of the soft gamma-ray sky across time or energy bands, forming a final image of 0.06$^\circ$ pixels.

Separate mosaics have been made for each set of GCDE observations in energy bands of 20--40 and 20--100~keV.   A zoom-in on the four mosaics (20--100~keV) centred on the location of IGR J17285-2922 is shown in Fig.~\ref{GCDE}, clearly illustrating that the source is visible only during observation C.  As a consequence of this, mosaics were also produced for individual revolutions in GCDE sets B and C in two energy bands (20--40 and 40--100~keV).   The source was not statistically significant in ScW images for any of the sets of observations.  Simultaneous images from the JEM-X instrument \citep{JEMX} were processed for revolutions 119 and 120 in the 3--20~keV energy range, but no detection was obtained.  Comparing the extrapolated ISGRI power law spectrum to the JEM-X sensitivity curve, we might expect a weak detection in this energy range.  However, we note that the region around IGR~J17285--2922 is heavily affected by systematic image artefacts, probably deriving from GX3+1, $\sim$5$^{\circ}$ away. Thus, the sensitivity we attain is far poorer than the statstical limit.  3$\sigma$ upper limits for the source flux, obtained from the JEM-X observations, are consistent with this scenario and are plotted in Fig.~\ref{sp150}.  The angular resolution ($\sim$degree) of the spectrometer, SPI \citep{SPI}, is unfortunately too coarse to obtain a clear detection of the source, due to close proximity of other bright sources.

\begin{figure}
\begin{center}
\includegraphics[totalheight=8.3cm,angle=-90]{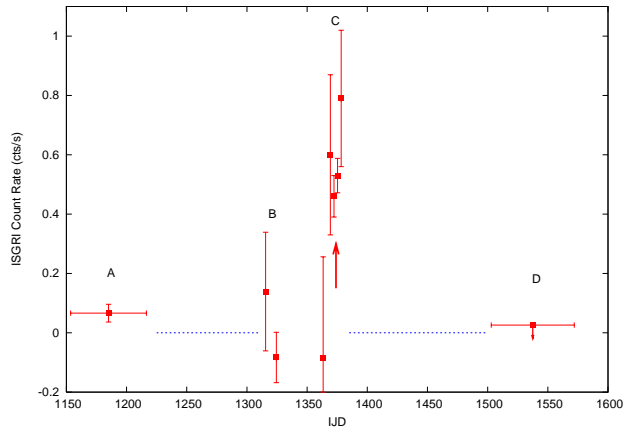}
\caption{\label{revlc} {Survey light curve for IGR~J17285--2922 (20--40~keV energy band).  Flux values are averaged over each revolution ($\sim$3 days) for observation sets B and C and averaged over the whole GCDE observation for A and D.  The dashed lines represent periods when the source could not be observed by IBIS and the single arrow indicates the revolutions (119 and 120) used for timing and spectral analysis.}}
\end{center}
\end{figure}

\subsection{ISGRI Timing Analysis}
Figure~\ref{revlc} summarises the behaviour of IGR~J17285--2922 throughout all of the GCDE observations. The source shows an outburst that started around IJD~1365, continuing for two weeks.  Unfortunately, the source then passed out of the IBIS field of view and was not observed again by {\it INTEGRAL} until four months later, by which time the source had returned to an undetectable quiescent state.  

\begin{figure}
\begin{center}
\includegraphics[totalheight=7cm, angle=-90]{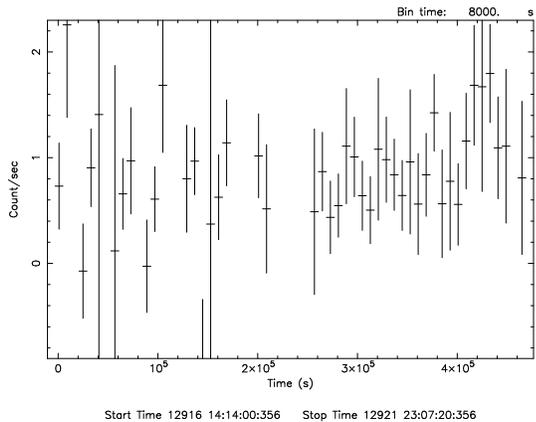}
\caption{\label{lc2000} {ISGRI light curve of IGR~J17285--2922 for revolutions 119 and 120 binned over 8000s (20--40~keV).  Time starts from the beginning of revolution 119, IJD=1372, where IJD~=~MJD~-~51544.}}
\end{center}
\end{figure}

For timing analysis, 193~ScWs from revolutions 119 and 120 were used,
of total duration 347~ksec.  In revolutions 118 and 121, the source
was mainly on the edge of the IBIS field of view as a consequence of
the dithering pattern of the telescope during the deep exposures.
Light curve information was extracted for all bright sources in the
field of view using OSA v4.1.  The software enables ScW and sub-ScW fluxes to be obtained for the source using the Pixel Illumination Function (PIF).  The PIF gives the relative illumination of each pixel by photons, from a given sky position, shining through the coded mask.  Events were selected with PIF=1 (i.e. the whole pixel was open to the source position), thereby minimising systematic effects over the whole detector plane due to photons that could not have originated from the source.  Fig.~\ref{lc2000} shows the 20--40~keV light curve using 8000s bins for data within 10$^\circ$off-axis. The average ISGRI source flux in 20--40~keV is 0.53~ISGRI c~s$^{-1}$.

\subsection{ISGRI Spectral Analysis}
\begin{figure}[h]
\begin{center}
\includegraphics[totalheight=7.5cm,angle=-90]{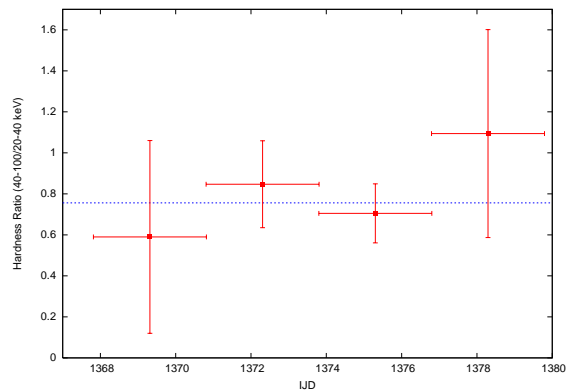}
\caption{\label{hr} { Flux$_{40-100}$/Flux$_{20-40}$ hardness ratio for revolutions 118 to 121 (bins $\sim$3 days).  Dotted line shows average hardness for GCDE C. }}
\end{center}
\end{figure}

In order to investigate if the source exhibits any spectral evolution, simple hardness ratios (F$_{40-100}$/F$_{20-40}$) were calculated for the four revolutions at the start of the source activity (118--121) using fluxes extracted directly from revolution mosaics.  Fig.~\ref{hr} plots the source hardness compared to the average value from fluxes extracted from GCDE mosaic C.  We note that the source did not show any significant change in hardness during our observations.

\begin{figure}
\begin{center}
\includegraphics[totalheight=6.1cm,angle=0]{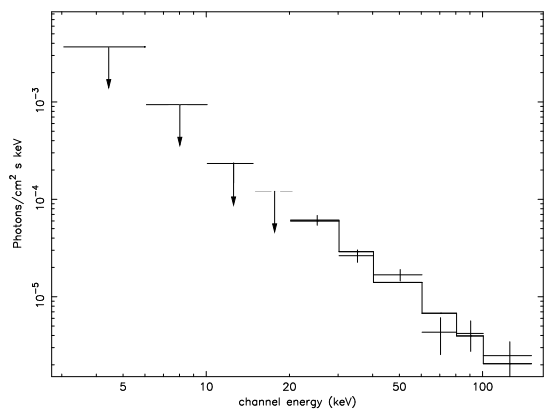}
\caption{\label{sp150} {20--150~keV ISGRI spectrum of IGR~J17285--2922 with simple power law model and 3--20~keV JEM-X 3$\sigma$ upper limits.}}
\end{center}
\end{figure}

 Fluxes for spectral analysis were extracted from fine-energy band mosaics of revolutions 119 and 120 added together.  Analysis was performed on 20--150~keV data in Xspec v11.3.1 with a systematic error of 5\%.  A simple power law model with a photon index $\Gamma = 2.1 \pm 0.17$ gives the best fit (Fig.~\ref{sp150}).  The flux in the 20--150~keV energy band is 1.1$\times$10$^{-10}$~erg~cm$^{-2}$~s$^{-1}$ corresponding to a luminosity estimate of 1$\times$10$^{36}$ erg~s$^{-1}$, assuming the source is at Galactic Centre distance of 8.5~kpc. We also used Comptonised (CompTT) and cut-off power law models \cite{comptt}, but statistics in the data were not good enough to constrain these models.  

\section{Discussion}
The spectral characteristics of the source and its location in the Galactic Bulge are consistent with IGR~J17285--2922 being a LMXB undergoing transient activity. 

Type I X-ray bursts are observed uniquely in neutron star (NS) systems
and explained as thermonuclear burning of material on the surface of the star.
Burst profiles typically display short rise times ($\leq$10~s) and
longer, exponential-type decays ($\geq$10~s--minutes) with spectra
described by blackbody radiation with a temperature of 2--3~keV (see Lewin et al. 1993 for
a review).  IGR~J17285--2922 has a low persistent luminosity of
$\sim$10$^{36}$ erg~s$^{-1}$ and, if a NS system, we might
expect it to display Eddington-limited bursts with a bolometric peak
flux of $\sim$10$^{39}$
erg~s$^{-1}$ and exponential decay times of 100--1000~s (Bildsten 1995).
We can work out the expected burst rate assuming a ratio of the total     
accretion energy released between bursts and burst fluence to be 100
\citep{Para}.  Assuming a typical burst fluence of
$\sim$10$^{39}$ erg gives an average time between bursts  
of 100~ksec for this source.  ISGRI has on-source exposure totalling over 300~ksec, giving a 3\% probability of no burst occuring during the observations.
For JEM-X (sensitive over 3--20~keV), although more capable of detecting bursts, the probability of no bursts occuring during
observations increases to 67\% due to reduced
source exposure ($\sim$40~ksec) as a result of the small field of view.   Therefore, it is rather likely we would miss all bursts with JEM-X.  For a
typical burst, we estimate that the peak luminosity should be at least 10$^{37}$ erg~s$^{-1}$ in
the 20--40~keV band.  Based on ISGRI sensitivity curves, this will
result in a significant ($\geq$9$\sigma$) detection in the 20--40 keV band for a 100~s
long burst, but a marginal ($\sim$3$\sigma$) detection for shorter
bursts of 10~s.  Although we fail to detect bursts with ISGRI 
(Fig.~\ref{lc2000}) and JEM-X (Sec. 2), this does not constrain our understanding of the nature of the compact object.

The 20--150 keV spectrum of the source is described by a power law with photon index $\Gamma$ = 2.1, thus showing evidence for the hard tail exhibited by black holes (BH).  For a BH in the low hard state (LHS), the spectrum is typically modelled by a power law with a high energy cut off, as a result of Comptonised emission from the inner disk \citep{BHBook}.  A good fit could not be obtained for such a model.  If this source is a BH in the LHS it may be that the spectral statistics are not good enough to constrain a value for the cut off, or the cut off is in excess of 100~keV.  In the high soft state (HSS), emission is dominated by a thermal disk component below $\sim$10~keV, no evidence of which is seen in JEM-X, see Sec. 2.  The 3--20~keV upper-limits are consistent with the source either being a BH/LMXB in a LHS \citep{BHBook}, or a member of a new class of obscured objects characterised by a lack of soft X-ray emission, interpreted as a Compton thick envelope enshrouding a HMXB \citep{4848}.

The period of activity (at least 2 weeks) of IGR~J17285--2922 detected by IBIS, could be the start of a stronger burst comparable to the transient periods of X-ray Novae, lasting of order of weeks or months and recurrance times of several years, see Tanaka \& Shibazaki \citep{BHB} for full review.  Historically, the vast majority of BHCs have been identified during X-ray Nova events including the {\it INTEGRAL} observed sources  XTE~1720--318 and XTE 1550--564 \citep{XTE, 1550}. Both of these sources have been observed, at some stage, to be accreting at sub-Eddington luminosities during non-quiescent states.  XTE~1720--318 was observed by INTEGRAL in outburst in Jan 2003 exhibiting a spectral state transition from HSS to LHS \citep{XTE}.  In the LHS, the 20--200 keV spectrum of XTE~1720--318 has been fit using a power law $\Gamma$=1.9, comparable with IGR~J17285--2922.   Peak 20--200 keV luminosity was sub-Eddington and 7 times more luminous than the average 20--150~keV luminosity of IGR~J17285--2922. Although Fig. \ref{hr} does not imply any spectral evolution during the observation of IGR~J17285--2922, the source is showing a similar relative hardness as XTE~1720--318 at the start of its state transistion.

At present, we are not able to constrain the nature of IGR~J17285--2922 from ISGRI data alone.  Although the luminosity is more consistent with NS systems, the lack of Type I X-ray bursts and the relative hardness of the spectrum suggests that the system may be harbouring a BH.  More observational data is needed to study this object further.  We will have to wait until the source outbursts again at which time a detection with {\it XMM-Newton} or {\it Chandra} would help improve the accuracy of the source location, crucial in the search for longer wavelength counterparts.  A detection in quiescence would point to the system containing a NS, as BH systems in quiescence are extremely faint, $\sim$10$^{31}$~erg~s$^{-1}$ \citep{BHBook}.  We hope to gather more observations of this system from the continuing survey campaign of {\it INTEGRAL}.
\begin{acknowledgements}
This work is based on observations with {\it INTEGRAL}, an ESA project with instruments and science data centre funded by ESA member states (especially the PI countries: Denmark, France, Germany, Italy, Switzerland, Spain), Czech Republic and Poland and with the participation of Russia and the USA.  Funding in the UK and for SES's position at the ISDC is provided by PPARC. 
\end{acknowledgements}

\end{document}